==================

\documentstyle[12pt]{article}

\begin{document}


\def\cc{\gamma\gamma\rightarrow W^+_{L} W^-_{L} }
\def\nc{\gamma\gamma\rightarrow Z_L Z_L}
\def\wl{W^{\pm}_{L}}
\def\zl{Z_L}
\def\sbp{SU(2)_L\times SU(2)_R \rightarrow SU(2)_{L+R}}
\def\gs{SU(2)_L\times SU(2)_R }
\def\ee{e^{+}e^{-} }
\def\phph{\gamma \gamma  }
\def\im{M_{WW}}

\begin{titlepage}
\pagestyle{empty}
\title{\bf Study of $ \cc \;$ and $\nc \;$ reactions with Chiral Lagrangians }

\author{
{\bf Mar!a Herrero}\thanks{
e--mail: herrero@vm1.sdi.uam.es}\ \ \  {\bf and Ester Ruiz-Morales}\thanks{
e--mail: meruiz@vm1.sdi.uam.es} \\[2mm]
Departamento de F!sica Te"rica\\
Universidad Aut"noma de Madrid\\
Cantoblanco\ \ \ E--28049\ \ Madrid ,\ \ Spain  }

\date{}
\maketitle
\def\baselinestretch{1.15}
\begin{abstract}
\noindent
We analyze the effects of a strongly interacting symmetry breaking
sector  of the Standard Model in $\cc \;$ and $\nc \;$ reactions at TeV
energies by using Chiral Lagrangians and Chiral Perturbation Theory.
We find significant deviations from the Standard Model predictions
for the differential cross sections at high
invariant mass of the gauge bosons pair. We study the experimental
signals that could be obtained in a high energy and high luminosity
dedicated $\gamma\gamma$ collider and estimate the sensitivity that such
experiments could reach to the values of the effective lagrangian parameters.

\end{abstract}

\vskip-15.5cm
\rightline{ FTUAM92/22}
\rightline{ August 1992}
\vskip3in

\end{titlepage}

\newpage
\def\baselinestretch{1}

\section{Introduction}

The interest of applying Chiral Lagrangians and Chiral
Perturbation Theory (ChPT) to the study of the
self-interactions of the longitudinal gauge bosons $\wl\;\;$ and $\zl\;\;$
at and below TeV energies has long been discussed in
literature [1]. The outcome of this discussion is that Chiral
Lagrangians are ideally suited for studying, with complete generality
and in the most economical way, the properties of the
symmetry breaking sector (SBS) in the hypothetical case that it is
strongly interacting (SISBS).This is so
because the only physical input they use is the symmetry breaking
pattern which must at least contain $\sbp \;$ if one wants to recover the
global $\gs\;$ invariance of the scalar sector of
the Standard Model (SM) when the U(1)$_Y$
coupling is set equal to zero, and because it only uses the so far
confirmed physical degrees of freedom of the SBS, namely the $\wl\;\;$
 and $\zl\;\;$.

The Chiral Lagrangian (ChL) is organized as an expansion in terms of
derivatives and external fields, and the different possibilities for
the SBS are just reflected in the different values for the parameters $L_i$'s
appearing in it [2]. Besides, ChPT provides us with a procedure
for expanding  observables in powers of $\displaystyle\frac{p^2}{(4\pi v)^2}$,
where  $4\pi v$ ($v$=246 GeV) is the scale that normalizes the
contribution of higher-dimensional operators and governs the size of
the $\wl\;\;$ and Z$_L$ self-interactions. Since   $4\pi v \simeq $ 3 TeV,
this expansion is  a very good one at energies well below  3 TeV.

Several studies have already been done on the sensitivity
to the values of the $L_i$ parameters that could be reached in the present
and future experiments, and to conclude at which level
one will be able to discriminate amongst the different symmetry breaking
alternatives from an experimental measure of those coefficients.
Most of the analyses so far have been dedicated to the process of
V$_L$ V$_L$ fusion ($V_L = \wl\;\;$ or $Z_L$) at the pp colliders LHC and
SSC, the reason being obviously that the reaction $V_L V_L
\rightarrow V_L V_L$ is the genuine one for testing the $V_L$'s
self-interactions in exactly the same way as the $\pi\pi \rightarrow
\pi\pi$ reaction does for the $\pi$'s self-interactions [3].
The sensitivity to the chiral parameters $L_1$ and $L_2$ that will be
reached in LHC and SSC through the $V_L V_L$ fusion processes has
been studied in [4]. More recently, ChL's have also
been applied to the process $q\bar q \rightarrow V_LV_L$
at LHC and SSC  at tree level [5], at one
loop [6], and  in the leading log approximation [7]. The study of
these processes will allow to constrain the numerical value of
$L_9$. Finally, $\ee\;$ colliders have also been used to test the chiral
parameters [8-11]. Given the fantastic accuracy that is being achieved at
LEPI, a study of the radiative corrections to the standard
parameters for LEP physics $\Delta r$, $\Delta\kappa$ and
$\Delta\rho$ by means of the ChL has allowed to constrain
quite significantly the value of $L_{10}$ [9]. On the other hand,
a tree level study in ChPT of the reaction $\ee\;\rightarrow W^+W^-$
first in [10] and a complete study in ChPT to one loop  later in [11]
have allowed to search for
the expected sensitivities of $L_9$ and $L_{10}$ at LEPII energies and at
the proposed $\ee\;$ colliders at TeV energies.

In this letter, we propose an alternative way of testing the SISBS
hypothesis, by means of the $\cc \;$ and $\nc \;$ reactions.
The interest of this analysis is two--fold. On one hand, the above
reactions complement nicely the study of SISBS with Chiral Lagrangians,
as the charched channel is sensitive to the $L_9$ and $L_{10}$ parameters
which do not contribute to $V_L V_L \rightarrow V_L V_L$ reactions
and the neutral channel is sensitive to the chiral loops.
On the other hand, the proposal of a dedicated $\phph\;$
collider up to TeV energies [12], where real photons are obtained from
backscattering of laser beams off high energy electrons,
opens the possibility of studying
these reactions in a very clean and almost background free way [13].
As we will comment later on, the peculiarities of a dedicated
$\phph\;$ collider
make this experiment optimal for our purpose, since one could reach the energy
region of 0.5 TeV $\leq \surd s \leq $ 1 TeV, where the effects of $V_L$'s
strong interactions are expected to show up, with high $\phph\;$ luminosities.

The use of $\phph\;$ collisions
at TeV energies will be the analogous one to the use of
$\phph\;$ $\rightarrow \pi \pi$ reactions at GeV energies.
As many studies have shown so far, the latter
has turned out to be a quite successful and fruitful laboratory to study
the effects of strong $\pi \pi$ rescattering in a very clean way.
We show in this work  that the study of $\cc \;$ and $\nc \;$ reactions in a
dedicated
$\phph\;$ collider will certainly provide useful information on the $\wl\;\;$
and
$\zl\;\;$ self-interactions at energies ranging from threshold to about 1 TeV.
In particular, we will see that the process $\cc \;$ will allow to constrain
quite significantly the values of the chiral parameters $L_9$
and $L_{10}$. Finally, we comment on what in principle could be a
 very promising signal of a SISBS.
It refers to the measure of the total and differential
cross section for the $\nc \;$ process. The interesting point is that, in
the SM, the cross section for this process vanishes
at tree level, and the only contributions come from one loop
diagrams that are of order $e^8$.
In contrast, if the SBS is strongly interacting and the
treatment by means of a ChL applies, the cross
section for $\nc \;$ will be of "enhanced electroweak strength"
due, essentially, to the
typical effect of strong rescattering of the $\zl \zl\;\;$ final
pair that translates into an overall contribution of the order of
$e^4$. Unfortunately, as we will show in this work, this potential
effect will not be seen in the $\phph$ dedicated colliders
with the luminosity considered here because of
the lack of statistics.

\section{Chiral Lagrangians and $T(\gamma \gamma \rightarrow V_{L} V_{L})$
 amplitudes}

The  effective ChL we
will work with  is based on the symmetry breaking pattern of the global chiral
symmetry down to its diagonal subgroup: $\sbp \;$. We will consider only
effective theories for which the custodial symmetry
$ SU(2)_{V}=SU(2)_{L+R}$ is
exactly preserved in the SBS up to the explicit breaking due to the
gauging of the $U(1)_Y$ symmetry. The effective lagrangian consists
of an infinite expansion of terms with increasing number of
derivatives of the Goldstone boson fields associated to the chiral
symmetry breaking $w^\pm$ and $z$, and/or external fields, namely the
$A_\mu$, $W^\pm_\mu$ and $Z_\mu$ gauge fields. The Goldstone boson
fields, that will eventually become the longitudinal components of
the $W^\pm$ and $Z$ fields, are parametrized by means of an unitary
matrix field $U$ belonging to que quotient space $\gs\;$/$SU(2)_V$:
\begin{equation}
 U=\exp (i\sigma^a w^a / v) , \ \ \ \  a=1,2,3.
\end{equation}
The local $SU(2)\times
U(1)_Y$ symmetry of the SM is implemented by replacing
the derivatives of the $U$ field by the covariant derivatives defined
as:
\begin{equation}
D_{\mu} U = \partial_{\mu} U + L_{\mu} U - U R_{\mu}
\end{equation}
where $L_\mu$ and $R_\mu$ are the left and right handed gauge fields:
\begin{equation}
L_{\mu} = i g \sigma^a W_{\mu}^{a}/2, \hspace{1.5cm}
 R_{\mu}= i g' \sigma^3 B_{\mu} /2.
\end{equation}
The complete effective lagrangian for the scalar sector and the gauge
boson sector of the SM that complies with all the symmetries of the
SM, and that includes terms up to ${\cal O}(p^4)$ is given by:
\begin{eqnarray}
{\cal L} & = & {\cal L}^{(2)} + {\cal L}^{(4)} +{\cal L}_{G} + {\cal L}_{GF}
+ {\cal L}_{FP} \\
{\cal L}^{(2)} & = & \frac{v^2}{4} Tr D_{\mu} U D^{\mu} U^\dagger\\
{\cal L}^{(4)} & = & L_{1}  Tr D_{\mu} U D^{\mu} U^\dagger  Tr D_{\nu} U
D^{\nu} U^\dagger  \nonumber\\
& & + L_{2} Tr D_{\mu} U D_{\nu} U^\dagger
Tr D^{\mu} U D^{\nu} U^\dagger \nonumber \\
& & - i L_{9} Tr ( F^{\mu \nu}_R D_{\mu} U D_{\nu} U^\dagger  +
F^{\mu \nu}_L D_{\mu} U D_{\nu} U^\dagger) \nonumber\\
& & + L_{10} Tr U^\dagger  F^{\mu \nu}_R U  F_{L \mu \nu} \\
{\cal L}_G & = & - \frac{1}{2 g^2} Tr  F_{L \mu \nu}  F^{\mu \nu}_L
- \frac{1}{2 g^{'2}} Tr  F_{R \mu \nu}  F^{\mu \nu}_R
\end{eqnarray}
with $F^{\mu \nu}_{L,R}$ being the field strengths associated to the
left and right gauge fields:
\begin{equation}
i F_{\mu \nu}^L = \partial_\mu L_\nu - \partial_\nu L_\mu -
[L_\mu,L_\nu] \; \;\;\;\;\;
i F_{\mu \nu}^R = \partial_\mu R_\nu - \partial_\nu R_\mu.
\end{equation}
${\cal L}_{GF}$ is the gauge-fixing term:
\begin{equation}
{\cal L}_{GF} =  - \frac{1}{2} \{ F^2_{\gamma} + F^2_{Z} + 2 F_{+}
F_{-} \}
\end{equation}
where:
\begin{eqnarray}
F_{\pm} & = & \frac{1}{\surd\xi} ( \partial_{\mu} W^{\mu \pm} - M_{W} \xi
w^{\pm}) \nonumber\\
F_{Z} & = & \frac{1}{\surd\xi} ( \partial_{\mu} Z^{\mu } - M_{Z} \xi z)\; ;
\hspace{1cm} F_{\gamma} = \frac{1}{\surd\xi} ( \partial_{\mu} A^{\mu}) .
\end{eqnarray}
We will work in the Landau gauge ($\xi=0$), where the ghosts and
Goldstone bosons decouple from each other and the computations are
simpler. It is also in this gauge where the goldstone bosons and
ghosts fields remain massless and therefore ${\cal L}_{GF}$ and
${\cal L}_{FP}$ will explicitely respect the
global $\gs\;$ Chiral Symmetry. The advantages of using this gauge
when dealing with a non--linear sigma model approach to the SM
were emphasized in [14], so we refer the reader to those references
for more details. Finally the ghost  term in the
Landau gauge is given by:
\begin{equation}
{\cal L}_{FP} = \partial_{\mu} \eta_0 \partial^{\mu} \chi_0 + 2 Tr [
\partial_{\mu} \eta ( \partial^{\mu} \chi + [L^{\mu},\chi])]
\end{equation}
where $\eta(x) = \vec{\sigma} \; \vec{\eta}(x)/2 ; \hspace{0.2cm}
\chi(x) = \vec{\sigma} \; \vec{\chi}(x)/2 \; $ and
$\eta^a(x) , \chi^a(x), a=1,2,3$ are the anticommuting scalar ghosts.

Once the complete Lagrangian is given, the derivation of the Feynman
rules needed for the computation of $\cc \;$ and $\nc \;$ amplitudes is
straightforward. At this point it is worth mentioning that these
Feynman rules are not the same as in the linear version of the
SM, the reason being obviously the non-linear realization
of the chiral symmetry in the ChL.

In order to compute the amplitudes for the reactions $\cc \;$ and $\nc \;$ we
have applied the equivalence theorem [3,15], that in this case states:
\begin{eqnarray}
T(\phph\rightarrow W^{+}_{L}W^{-}_{L})&=& - T(\phph\rightarrow w^{+}
w^{-}) + {\cal O}(\frac{M_{W}}{\sqrt{s}}) \nonumber\\
T(\phph\rightarrow Z_{L} Z_{L})&=& - T(\phph\rightarrow z z) +
 {\cal O}(\frac{M_{Z}}{\sqrt{s}})
\end{eqnarray}
so that the range of applicability will be restricted to high energies
as compared to the gauge boson masses.
As we will show below, for
the process considered here it turns out to be indeed a very good
approximation for energies larger than 400 GeV.

Let us now present
the results for the Goldstone boson amplitudes obtained from the
ChL to order $p^4$ in equations (4--11). We work to
the lowest order in the gauge coupling constant and to one loop in
the chiral expansion. This means that we keep only contributions
in the amplitudes of
order $e^2$ coming from the ChL to tree level, namely
from ${\cal L}^{(2)}$ and ${\cal L}^{(4)}$ ,
and use only ${\cal L}^{(2)}$ to compute the one loop contributions.
The contributions from the loops generated by ${\cal L}^{(4)}$
are of higher order in the
chiral expansion, i.e. of order $p^6$, and we will not consider them here.
It is important to emphasize that once we have replaced the external fields
by the Goldstone bosons and, whenever we work at order $e^2$ in the
amplitudes, the only one-loop diagrams from ${\cal L}^{(2)}$ that
contribute to our processes are those with just Goldstone bosons circulating
in the loops.

The amplitudes written down below are presented in the form:
\begin{equation}
 T = T^{(2)} + T^{(4)}\nonumber
\end{equation}
where we have separated explicitly the contributions from ${\cal L}^{(2)}$ at
tree level in  $T^{(2)}$ and the contributions from ${\cal L}^{(4)}$
at tree level plus those from ${\cal L}^{(2)}$ at one loop in $T^{(4)}$.

We get the following results for the various helicity amplitudes
$T_{\lambda_{1} \lambda_{2}}$ with $\lambda_{1,2}$ being the initial
photon helicities:
\begin{eqnarray}
\hspace{-0.5cm}T_{++}(\gamma\gamma\rightarrow W^{+}_{L} W^{-}_{L})&=&
T^{(4)}_{++}(\gamma\gamma\rightarrow W^{+}_{L} W^{-}_{L})
= -  16 \pi \alpha \frac{\hat{s}}{(4 \pi v)^2} \left[16 \pi^2 (L_9+
L_{10}) - \frac{1}{4}\right] \nonumber\\
 & & \\
T_{+-}(\gamma\gamma\rightarrow W^{+}_{L} W^{-}_{L})&=&
T^{(2)}_{+-}(\gamma \gamma \rightarrow W^{+}_{L} W^{-}_{L})
= -  8 \pi \alpha
\end {eqnarray}
and for the neutral channels:
\begin{eqnarray}
T_{++}(\gamma \gamma \rightarrow Z_{L} Z_{L}) & = &
 T^{(4)}_{++}(\gamma \gamma \rightarrow Z_{L} Z_{L})
 = - 8 \pi \alpha \frac{\hat{s}}{(4 \pi v)^2} \\
T_{+-}(\gamma \gamma \rightarrow Z_{L} Z_{L}) & = & 0
\end{eqnarray}
In the above expressions $\surd \hat{s}$ is the center of mass
energy of the $\phph\;$
$\rightarrow V_LV_L$ subprocess. The rest of the helicity
amplitudes are related to the above ones by $T_{--} = T_{++}$ and
$ T_{-+} = T_{+-} $.

Finally, from these expressions we get the following predictions in ChPT for
the total cross sections with unpolarized photons \footnote{
We have checked that these expresions coincide with the corresponding ones
for $\gamma \gamma \rightarrow \pi^+ \pi^-$ and
$\gamma \gamma \rightarrow \pi^o \pi^o$ in [16] once the chiral
limit is taken. We would like to emphasize here that a formal
connection between $V_L$ and $\pi$ physics may only be stablished if the
Landau gauge has been chosen in the former and the chiral limit is taken
in the latter.}:
\begin{eqnarray}
\hat{\sigma}^{ChPT}(\gamma \gamma \rightarrow W^{+}_{L} W^{-}_{L})
& = & \displaystyle\frac{2 \pi \alpha^2}{\hat{s}} \left\{1 + 4 \left[
\frac{\hat{s}}{(4 \pi v)^2}
\left(16 \pi^2 (L_9 + L_{10}) - \frac{1}{4}\right)\right]^2 \right\}\\
\hat{\sigma}^{ChPT}(\gamma \gamma \rightarrow Z_{L} Z_{L}) & = &
\frac{ \pi \alpha^2}{\hat{s}} \left[\frac{\hat{s}}{(4 \pi v)^2}\right]^2
\end{eqnarray}
At this point, it is worth commenting on some aspects of the above
results that we find interesting. First of all, the two
cross sections turn out to be finite in ChPT at one loop. This means
that they are fully predictable and do not need of the
renormalization procedure that is usually implemented when computing
in ChPT. In the case of the neutral channel, this finiteness is
explicitely shown in eq.(19), where potential terms of O($1/\epsilon$)
from dimensional regularization are absent. The finiteness of the
charged channel result is understood easily from eq.(18) by noticing
the particular combination of the chiral parameters appearing there,
namely $L_9 + L_{10}$, and recognizing it as a renormalization
group invariant quantity that implies  $L_9^{r}+L_{10}^{r} =
L_9+L_{10} $, with  $L_i^{r}$ being the renormalized parameters.

Another interesting aspect is  the result for the neutral channel being
independent of the chiral parameters $ L_{9}$ and $L_{10}$ and only dependent
on the dimensionful parameter $v$. In other words, the
cross section for $\nc \;$\ in ChPT does not depend on the precise
underlying dynamics governing the interactions amongst the longitudinal
gauge bosons. Thus, in principle we could use this channel to
isolate, and in consequence to test, the part of the $V_L$ self-
interactions that is universal, including chiral (Goldstone bosons) loops.
The result for the neutral channel is shown in Fig.1a.
In contrast, the charged channel does depend on the $L_9$ and $L_{10}$ chiral
parameters, so that in principle one could use this channel to test
the different alternatives for the SBS and thus to constrain their
numerical values when comparing with experimental data.

In order to study the differences between a strongly interacting SBS
and a weakly interacting SBS, we have chosen to compare the above
predictions in ChPT with the corresponding ones in the SM.
As can be seen in Fig.1b and Fig.1c, the total cross section for $\cc \;$
at energies ranging from 500 GeV to 1.5 TeV, is indeed quite
sensitive to the particular values of $L_9 + L_{10}$ with the raising of
the cross section in the high energy region becoming apparent for
values of $L_9$ + $L_{10}$ larger than $4/16\pi^2$.
An explicit computation of the cross section $\sigma(\cc \;$) in the
SM at tree level using the equivalence theorem approximation
and a comparison with the exact
result in the SM teaches us (see Fig.1c) that it is indeed a very good
approximation for energies above 400 GeV.

Finally, it should be mentioned that the cross sections predicted in ChPT,
eqs.(18) and (19) are not well behaved at high energies since the
corrections to the lowest order results grow with energy. This behaviour
produces as usual in ChPT a violation of perturbative unitarity in the
amplitudes at high energy. This problem is well known to happen in the
context of $\gamma\gamma \rightarrow \pi\pi $  reactions where many
implementations of unitarization have been proposed in the literature
[16,17].
The common assumption in these works is to relate the unitarization
of  $\gamma \gamma \rightarrow \pi\pi $ reactions with the corresponding one
for the final state $ \pi\pi $ interactions. Thus, the energy scale
at which unitarity is violated in  $\gamma\gamma \rightarrow \pi\pi $
is deduced directly from the corresponding one in
$\pi\pi \rightarrow \pi\pi $ scattering. A similar analysis for the
$\cc \;\; $ and $\nc\;\; $ amplitudes can be performed. The
conclusion is that the maximum allowed energy so that the results are
compatible with the unitarity bounds is $\surd s \simeq 1.5$ TeV.

Clearly, in order to make effective such a detailed analysis of the reactions
involving longitudinal gauge bosons in the final state one should be
able to discern experimentally longitudinal  from transverse modes.
This may be a hard task in pp supercolliders but we believe
it will be feasible in high precision experiments as for instance the
$\phph\;$
collider at TeV energies that we consider in this work.
In this concern, some strategies for measuring the polarization of
the final $ W_L$'s by studying the correlation between the $W$ pair
decay planes have been proposed in the literature in the context of
pp super-colliders [18]. We will not apply these techniques here
but prefer to postpone this kind of analysis until a more definite
proposal of a dedicated $\phph \;\;$ collider be done.
Instead of doing this, we have computed and compared the total and
differential cross sections in the SM at tree level for the three
different polarization states of the final $W^+ W^- $ pairs.
We get the following results for the total cross sections:
\begin{eqnarray}
\hat{\sigma}(\hat{s})_{LL} &=&
\frac{\pi \alpha^2 \beta}{\hat{s} r} \left[
\frac{2 ( r + 3 r^2 + r^3 + r^4)}{(1-r)^2} -
\frac{r^2 (2 + 3 r + 2 r^2 - r^3)}{(1-r)^2} \frac{1}{\beta}
\ln \left( \frac{1 + \beta}{1 - \beta} \right) \right] \\[2mm]
\hat{\sigma}(\hat{s})_{LT} &=&
\frac{\pi \alpha^2 \beta}{\hat{s} r} \left[
\frac{- 32 r^2 + 8 r^3 }{(1-r)^2} +
\frac{16 r^2 - 8 r^3 + 4 r^4}{(1-r)^2} \frac{1}{\beta}
\ln \left( \frac{1 + \beta}{1 - \beta} \right) \right] \\[2mm]
\hat{\sigma}(\hat{s})_{TT} &=&
\frac{\pi \alpha^2 \beta}{\hat{s} r} \left[
\frac{32 - 60 r + 52 r^2 - 16 r^3 + 4 r^4}{(1-r)^2} - \right.\nonumber \\
 & &\left. \frac{r^2 ( - 20 + 26 r - 14 r^2 + 2 r^3)}{(1-r)^2} \frac{1}{\beta}
\ln \left( \frac{1 + \beta}{1 - \beta} \right) \right]
\end{eqnarray}
where $\displaystyle r = \frac{4 M_W^2}{\hat{s}} $ and $\beta = \sqrt{1 - r} $.

The numerical results shown in Fig.2a, indicate clearly that the
transverse modes dominate by far the total cross section. For
instance, $\sigma_{TT}$ at $ \surd s = $ 1 TeV is almost three orders
of magnitude larger than $\sigma_{LL}$. As announced, this confirms
the absolute necessity of isolating the longitudinal from the
transverse modes in the proposed $\phph \;\; $ experiment.
One possibility we have found  will help in this task  is by
considering the differential cross section with respect to the
angular variables. For instance, one may compare the different
patterns in Fig.2b for the three combinations TT, LT and LL in the
variable $\cos \theta$, where $\theta$ is the angle of the $W$ in the
center of mass system. By applying a cut of $ | \cos \theta | \leq 0.8 $ we see
that the cross section for TT modes gets reduced in almost a two orders
of magnitude factor, whereas the cross section for LL modes remains
practically unchanged. Of course, a more realistic study in terms of
the leptonic and/or hadronic final states from the $ W$'s and $Z$'s
decays should be done in order to be able to reach a definite
conclusion.
For the rest of this work, we will assume that this separation
between the longitudinal and transverse modes can be done.

\section{Observable cross sections}

$\cc \;$ and $\nc \;$ reactions at TeV energies could, in principle, be studied
in $\ee\;$ colliders by means of virtual photon fusion processes.
However, the photon-photon luminosity is too small at high energies as
to produce a significant number of W pairs with large
invariant mass (0.5 $\leq \im \leq $ 1 TeV). Since it is in this
$\im\;$ range where the effects we are discussing are expected to show
up, they seem to be hardly detectable in this type of colliders.

We will analize instead the experimental signals that one would
obtain, in the case of a SISBS, through $\cc \;$ and $\nc \;$ reactions in a
dedicated $\phph\;$ collider, proposed by Ginzburg et al. [12]. They
showed how to transform a linear $ \ee \; $ or $ e^- e^- \;\;$
collider into a $\phph\;$  collider
with approximately the same energy and luminosity as the original $\ee\;$
collider. Each $\gamma$ beam is obtained by backward Compton
scattering of laser light focused on the e$^-$ beams, near the point
where the beams of scattered photons will finally collide.
When laser photons with energy $\omega_{o}$ scatter off electrons
with high energy $E$
at very small scattering angle, the beam of scattered photons
emerges along the incident e$^-$ direction with very small angular
spread, and an energy ($\omega$) spectrum given by
\footnote{ We have taken the approximation of zero
scattering angle $\alpha_{o}=0$ and negligible beam spread at the
focal spot $\rho \ll 1$. The conversion coefficient $\kappa$ is taken
here equal to 1.} [12]:
\begin{equation}
f(x,y) = \frac{2 (1+x)^2 ( 2 x^2 - 4 x y - 4 x^2 y + 4 y^2 + 4 x y^2
+ 3 x^2 y^2 - x^2 y^3)}{(1-y)^2 ( x (16 + 32 x + 18 x^2 + x^3) -
2 ( 8 + 20 x + 15 x^2 + 2 x^3 - x^4) \log(1+x))}
\end{equation}
\begin{equation}
x = \frac{4 E \omega_{o}}{m_e^2} \hspace{1cm} y = \frac{\omega}{E}
\nonumber
\end{equation}
The variable $x$ is related to the center-of-mass energy
of the Compton process by
$s_c = m_e^2 (x+1)$, and $y$ is the energy of the scattered photon
in units of E. For a given value of $x$, determined by the
experimental setup, there is an upper limit for the energy of the scattered
photons $y_{m}$:
\begin{equation}
 y \le y_m = \frac{x}{x+1}
\end{equation}
One would like to increase $x$ (i.e. $E$ or
$\omega_{o}$) in order to get values of $y_m$ as high as possible.
However, when $x > 2 + 2 \sqrt{2}$, some other processes besides
Compton scattering become important in the conversion region,
mainly $e^+ e^-$  pair production from a laser photon and a high
energy scattered photon. In order to avoid these background processes,
we have fixed $x$ to the value $2 + 2 \sqrt{2}$ which implies $y_m =$
0.828.
We have chosen, for definiteness, a $ \ee \;$ collider with  integrated
luminosity of $10$ fb$^{-1}$  and beam energies of 0.5 and 1 TeV
(the first case corresponds to the parameters of VLEPP [19].)
By taking $x=2 + 2 \sqrt{2}$, these energies will imply
using lasers with $\omega_o \simeq $ 0.3 and 0.15 eV respectively.

The total observable cross sections for $\cc \;$ and $\nc \;$
reactions in the collider described above are obtained from the
corresponding subprocess cross sections by convoluting them with the
two photons spectra. For the charged channel:
\begin{equation}
\sigma(\gamma \gamma \rightarrow W^{+}_{L} W^{-}_{L}) =
\int^{y_m}_{0} dy_1 \int^{y_m}_{0} dy_2 \hspace{0.2cm}
\left[ f(x,y_1) \;\cdot\; f(x,y_2) \right] \hspace{0.2cm}
\hat{\sigma}(\gamma \gamma \rightarrow W^{+}_{L} W^{-}_{L})
\end{equation}
where $\hat{\sigma}$ is given in ChPT in eq.(18) and in the SM in
eq.(20), and $f(x,y_i)$ is the spectrum of the $i$-photon given in
eq.(23).

The differential cross section with the invariant mass of the final
gauge bosons pair $M_{VV}$, is then given by:
\begin{equation}
\frac{d \sigma}{d M_{WW}} = \left[ \frac{d {\cal L}(z)}{d z} \right]
\frac{1}{\sqrt{s}}
\hat{\sigma}(\gamma \gamma \rightarrow W^{+}_{L} W^{-}_{L})
\end{equation}
where:
$z = \frac{M_{WW}}{\sqrt{s}} = \sqrt{\frac{\hat{s}}{s}} $; $\sqrt{s}$
is the total $e^+ e^-$ energy,  $\sqrt{\hat{s}}$ is the total
 $\phph $ energy and the spectral photon-photon luminosity is given by:
\begin{equation}
\frac{d {\cal L}(z)}{d z} = 2 \; z \int_{z^2/y_m}^{y_m} dy
\frac{f(x,z^2/y) \; f(x,y)}{y}.
\end{equation}
Similar expressions to eqs.(26) and (27) can be found for the neutral
channel.
Notice that the maximum value for $z$ is fixed to $y_m$, and
therefore, the maximum value of the final $VV$ pair invariant mass
$M_{VV}$ is 0.828 and 1.656 TeV for $\sqrt{s} =$ 1 and 2 TeV
respectively.
The final numerical results for  $\displaystyle \frac{d \sigma}{d M_{WW}}$ and
$\displaystyle\frac{d \sigma}{d M_{ZZ}}$ are shown in Fig.3 and
Fig.4, respectively. Only
invariant mass values larger than 0.5 TeV are displayed for the
charged channel. In this
energy range, a perfect agreement between the exact tree level SM
result and the corresponding one by using the Equivalence Theorem is
found. The predictions for $\displaystyle\frac{d \sigma}{d M_{WW}}$ in ChPT
show
a sizeable enhancement over the SM result at high invariant mass
$M_{VV}$ values, mainly for the $\sqrt{s}=2$ TeV $\ee$ collider.
We have chosen in the plots, for definiteness, two particular values
for $ L_{9}+L_{10}$, but the same analysis can be performed for any
other choice.

In order to study the sensitivity of these experiments to the values
of the Chiral parameters we have computed the number of $W^+_L W^-_L$
events that are expected in one year at a dedicated $ \phph \;$
collider with parameters $\sqrt{s} = $ 1 TeV and 2 TeV and integrated
one-year-luminosity of ${\cal L} = $10 fb$^{-1} $. We compare the
predicted number of events within an invariant mass range of
0.5 $\leq M_{WW} \leq M_{WW}^{max}$  for a given value of ($L_9 +
L_{10}$) , $N_{ChPT}(L_9 + L_{10})$, with the corresponding prediction
in the SM, $N_{SM}$, that we take always as a reference number
characterizing a weakly interacting SBS. For each chosen value of ($L_9 +
L_{10}$), we define our signal as the difference:
\begin{equation}
\Delta N = N_{ChPT}(L_9 + L_{10}) - N_{SM},
\end{equation}
which represents the excess of $W^+_L W^-_L$ events over the SM
prediction at large invariant mass $M_{WW} \geq $ 0.5 TeV.
For the upper limit of the invariant mass  we take the maximum value allowed
by unitarity constraints $ M_{WW}^{max} = $ 1.5 TeV in the $\sqrt{s}=
$ 2 TeV case and  the maximum imposed by $y_m \;$, $ M_{WW}^{max} =
y_m  \sqrt{s} = $ 0.828 TeV in the $\sqrt{s} =$ 1 TeV case.

In order to estimate the statistical significance of the effect we are
looking for, we have also evaluated the variable F, defined as
F = $\Delta N / \sqrt{N_{SM}} $, as a function of $(L_9 + L_{10})$. This
variable is an estimator of the number of sigmas measuring the
confidence level of the hypothesis that $(L_9 + L_{10}) \neq 0$
produces a noticeable effect.
The results are given in Table 1, which show that this experiment will be
indeed very sensitive to the values of $(L_9 + L_{10}) \neq $ 0.
Thus, for instance, if a value for the statistical significance as
large as F = 2 is required, the values of $(L_9 + L_{10})$ that could
be tested in a dedicated $\phph \;$ collider are:
\begin{center}
$\displaystyle (L_9 + L_{10}) \leq - \frac{0.9}{16\pi^2} $ \hspace{0.5 cm} and
\hspace{0.5cm} $ \displaystyle(L_9 + L_{10}) \geq  \frac{1.5}{16\pi^2} $
\hspace{0.5cm}
for $\sqrt{s} =$ 2 TeV
\end{center}
and
\begin{center}
$\displaystyle (L_9 + L_{10}) \leq - \frac{2.4}{16\pi^2} $ \hspace{0.5 cm} and
\hspace{0.5cm} $\displaystyle (L_9 + L_{10}) \geq  \frac{3.0}{16\pi^2} $
\hspace{0.5cm}
for $\sqrt{s} =$ 1 TeV
\end{center}

For the neutral channel, the corresponding number of events that are
predicted  in one-year running are of just $N_{ChPT} = 1$ at $\sqrt{s}
= 1$ TeV and  $N_{ChPT} = 6$ at $\sqrt{s}= 2$ TeV. Unfortunately, the
lack of statistics will not allow to make any detailed analysis of
the contributions of the Chiral loops in this channel.

\section{Conclusions}

The proposed dedicated $\phph $ colliders at TeV energies, where real
photons are obtained from backward Compton scattering of laser beams
off high energy electrons, will offer the possibility of studying the
reactions $\cc\; $ and $\nc\; $ in the high invariant mass region.
We have profited from this advantage to study their potentiality in
testing the Symmetry Breaking Sector of the SM and the longitudinal
gauge bosons self-interactions.

Our study by means of a Chiral Lagrangian approach of these reactions
show that the neutral channel will probe the pure strong $ V_L V_L $
 rescattering effects (i.e. Chiral loops) while the charged channel
will also probe the numerical values of the Chiral parameters $ L_9$
and $L_{10}$. The charged channel cross section turns out to depend
on the renormalization group invariant combination $(L_9 + L_{10}) $
and therefore is a fully predictable quantity in ChPT.

As a result of our analysis on the expected sensitivities to
$(L_9 + L_{10}) $ we have been able to place bounds on their minimum
positive and their maximum negative values that will be significantly
tested in these experiments.

Of course, a definite conclusion can not be drawn until a complete
analysis of the signal and backgrounds including the final decays of the
$Z_L$'s and $W_L$'s be done.

\section*{Acknowledgements}

We are indebted to A.Dobado for useful and interesting discussions.
This work has been partially supported by the Ministerio de Educaci"n
y Ciencia (Spain) (CICYT AEN90-272).

\newpage

\section*{References}
\begin{description}
\item[{[1]}] \hspace{1.7mm}
     A.Dobado and M.J.Herrero, Phys.Lett.{\bf B228}(1989)495;
     {\bf B233}(1989)505\\
     J.Donoghue and C.Ramirez, Phys.Lett.{\bf B234}(1990)361\\
     For a recent review on this subject see S.Sint, Diplomarbeit
     Universitat Hamburg 1991
\item[{[2]}]\hspace{1.7mm}
     S.Weinberg, Physica {\bf A96}(1979)327\\
     J.Gasser and H.Leutwyler, Ann.Phys.(N.Y.)158(1984)142; Nucl.Phys.
     {\bf B250}(1985)465
\item[{[3]}]\hspace{1.7mm}
     M.S.Chanowitz and M.K.Gaillard, Nucl.Phys.{\bf B261}(1985)379\\
     O.Cheyette and M.K.Gaillard, Phys.Lett.{\bf B197}(1987)205
\item[{[4]}]\hspace{1.7mm}
     A.Dobado, M.J.Herrero and J.Terron, Z.Phys.{\bf C50}(1991)205;
     Z.Phys.{\bf C50}(1991)465\\
     S.Dawson and G.Valencia, Nucl.Phys.{\bf B352}(1991)27\\
     H.Veltman and M.Veltman, Acta Phys.Pol.{\bf B22}(1991)669
\item[{[5]}]\hspace{1.7mm}
     A.Falk, M.Luke and E.Simmons, Nucl.Phys.{\bf B365}(1991)523
\item[{[6]}]\hspace{1.7mm}
     J.Bagger, S.Dawson and G.Valencia, Preprint Fermilab-Pub-92/75-T
\item[{[7]}]\hspace{1.7mm}
     A.Dobado and M.Urdiales, Preprint FT/UCM/11/92, to appear in Phys.Lett.B.
\item[{[8]}]\hspace{1.7mm}
     A.Dobado and M.Herrero, Phys.Lett.{\bf B233}(1989)505
\item[{[9]}]\hspace{1.7mm}
     M.Peskin and T.Takeuchi, Phys.Rev.Lett.{\bf 65}(1990)964\\
     B.Holdom and J.Terning, Phys.Lett.{\bf B247}(1990)88\\
     G.Altarelli and R.Barbieri, Phys.Lett.{\bf B253}(1991)161\\
     A.Dobado, D.Espriu and M.Herrero, Phys.Lett.{\bf B255}(1991)405\\
     M.Golden and L.Randall, Nucl.Phys.{\bf B361}(1991)
\item[{[10]}]
     B.Holdom,Phys.Lett.{\bf B258}(1991)156
\item[{[11]}]
     D.Espriu and M.Herrero, Nucl.Phys.{\bf B373}(1992)117
\item[{[12]}]
     I.F.Ginzburg et al., Nucl.Instrum.Meth.{\bf 205}(1983)47\\
     I.F.Ginzburg et al., Nucl.Instrum.Meth.{\bf 219}(1984)5
\item[{[13]}]
     I.F.Ginzburg et al.,Nucl.Phys.{\bf B228}(1983)285\\
     E.Yehudai, Preprint SLAC-Pub-5495/91
\item[{[14]}]
     T.Appelquist and C.Bernard, Phys.Rev.{\bf D22}(1980)200\\
     A.Longhitano, Nucl.Phys.{\bf B188}(1981)118\\
     T.Appelquist, "Broken Gauge Theories and Effective Lagrangians"
     in Proceedings of The XI Scottish Universities Summer School in
     Physics, page 385, Ed. by K.Bowler and D.Sutherland.
\item[{[15]}]
     J.M.Cornwall, D.N.Levin and G.Tiktopoulos, Phys.Rev.{\bf D10},
     1145(1974)\\
     B.Lee, C.Quigg and H.Thacker, Phys.Rev.{\bf D16},1519(1977)\\
     G.J.Gounaris, R.Kogerler and H.Neufeld, Phys.Rev.{\bf D34},
     3257(1986)\\
     H.Veltman, Phys.Rev.{\bf D41},2294(1990)
\item[{[16]}]
     J.Bijnens and F.Cornet, Nucl.Phys.{\bf B296}(1988)557\\
     J.Donoghue, B.Holstein and Y.Lin, Phys.Rev.{\bf D37}(1988)2423
\item[{[17]}]
     D.Morgan, M.R.Pennington, Phys.Lett.{\bf B192}(1897)207\\
     R.Goble, R.Rosenfeld and J.L.Rosner, Phys.Rev.{\bf D39}(1989)3264\\
     P.Ko, Phys.Rev.{\bf D41}(1990)1531\\
     A.Dobado and J.Pelaez, Preprint FT/UCM/8/92
\item[{[18]}]
     M.Duncan, G.Kane and W.Repko, Nucl.Phys.{\bf B272}(1986)517\\
     G.Kane and C.P.Yuan, Phys.Rev.{\bf D40}(1989)2231
\item[{[19]}]
     V.E.Balakin and A.N.Skrinsky, Preprint INP-81-129 (Novosibirsk,
     1981)\\
     A.N.Skrinsky, Ups.Fiz.Nauk.{\bf 138}(1982)3\\
     Review of Particle Properties, Phys.Lett.{\bf B239}(April 1990)

\end{description}

\newpage
\section*{Figure Captions}
\begin{description}
\item[Fig.1] Predictions for the subprocess $\gamma \gamma
\rightarrow V_L V_L \;$ cross sections to order $\alpha^2$
 as a function of the $VV$ pair invariant mass $M_{VV}$.\\
{\bf (1a)} Chiral prediction for $\nc \;$ reaction. The SM cross section
is zero at tree level.\\
{\bf (1b)} Behaviour of the $\cc\; $ cross section in the
high $M_{WW}$ region. The Chiral prediction is shown for values of
 $L_9+L_{10} $ equal to $1/16\pi^2$ (short-dashed line),$ 2/16\pi^2$
 (dashed line) and $4/16\pi^2$ (long-dashed line),
and compared with the SM result.\\
{\bf (1c)} Cross sections for the $\cc \;$ reaction. The SM prediction at
tree level (solid line) is compared with the result obtained using
the Equivalence Theorem approximation (dotts) and the complete Chiral
prediction for $L_9+L_{10}=4/16\pi^2$ (dashed line).\\
\item[Fig.2] Contributions of the different polarization states (LL,
LT and TT) of the final $W^+ W^-$ pair to the SM cross sections in
$\cc\;$ reactions.\\
{\bf (2a)} Total cross section as a function of $M_{WW}$.\\
{\bf (2b)} Differential cross section as a function of cos$\theta$,
where $\theta$ is the scattering angle of the final W in the CM-system.
\item[Fig.3]  Differential cross section as a function of $M_{WW} $  for
$ W^+_L W^-_L $ pair production in the dedicated $\phph \;$ collider
described in the text.\\
{\bf (3a)}
 Results for an original $ ee $ collider with a
CM energy of 2 TeV. The results obtained for the SM
in the region $0.5 \leq M_{WW} \leq M_{WW}^{max} $ are compared with
the Chiral prediction for $L_9+L_{10}=4/16\pi^2$  (long-dashed line) and
$L_9+L_{10}=-2/16\pi^2$  (short-dashed line). The points correspond
to the predictions with the Equivalence Theorem approximation.\\
{\bf (3b)} The same as 3.a for an $ee$ collider of 1 TeV CM energy.
\item[Fig.4] Differential cross section as a function of $M_{ZZ} $  for
$ Z_L Z_L $ pair production in the dedicated $\phph \;$ collider
described in the text. The results are shown for ee colliders with
 CM energies of 1 (short-dashed line) and 2 (long-dashed line) TeV.\\

\end{description}

\newpage

\begin{table}[t]
\caption[]{ }
\vspace{1cm}
\centering
\begin{tabular}{||rl|c|r||rl|c|r||}  \hline\hline
\multicolumn{4}{||c||}{ $\sqrt{s} =$ 1 TeV } &
\multicolumn{4}{|c||}{ $\sqrt{s} =$ 2 TeV } \\ \hline
\multicolumn{4}{||c||}{ $N_{SM} = 1421 $ } &
\multicolumn{4}{|c||}{ $N_{SM} = 1416 $ } \\ \hline
\multicolumn{2}{||c|}{$(L_9 + L_{10})*16\pi^2 $ } & $ N_{ChPT} $ &  F &
\multicolumn{2}{|c|}{$(L_9 + L_{10})*16\pi^2 $ } & $ N_{ChPT} $ & F \\ \hline
\hspace{5mm}-&4 & 1603 & 4.82 &\hspace{5mm} -&4 & 2216 & 21.26 \\ \hline
\hspace{5mm}-&3 & 1529 & 2.88 &\hspace{5mm} -&3 & 1890 & 12.60 \\ \hline
\hspace{5mm}-&2 & 1476 & 1.46 &\hspace{5mm} -&2 & 1651 & 6.26 \\ \hline
\hspace{5mm}-&1 & 1442 & 0.55 &\hspace{5mm} -&1 & 1499 & 2.22 \\ \hline
   &0 & 1427 & 0.16 & &0 & 1434 & 0.49 \\ \hline
   &1 & 1432 & 0.29 & &1 & 1456 & 1.07 \\ \hline
   &2 & 1456 & 0.94 & &2 & 1564 & 3.95 \\ \hline
   &3 & 1500 & 2.10 & &3 & 1760 & 9.15 \\ \hline
   &4 & 1563 & 3.78 & &4 & 2042 & 16.64 \\ \hline\hline
\end{tabular}
\end{table}

\end{document}